\documentclass[11pt]{article}
\newcommand{\Comment}[1]{{}}
\usepackage{amsfonts,amsthm,amsmath,amssymb,slashed}
\usepackage[textwidth = 430 pt, textheight = 630 pt]{geometry}
\usepackage{color}

\Comment{\usepackage{color}
\definecolor{MyDarkBlue}{rgb}{0.15,0.15,0.45}
\usepackage[linktocpage=true]{hyperref}
\hypersetup{
colorlinks=true,
citecolor=MyDarkBlue,
linkcolor=MyDarkBlue,
urlcolor=MyDarkBlue,
pdfauthor={Horatiu Nastase and Jacob Sonnenschein},
pdftitle={Euler fluid as a gauge theory, and an action for the Euler fluid},
pdfsubject={hep-th}
}

\usepackage[numbers,sort&compress]{natbib}
\usepackage{hypernat}}
\usepackage{graphicx}
\usepackage{cite}

\newcommand\ignore[1]{}
\def\one{{\,\hbox{1\kern-.8mm l}}}

\def\a{\alpha}\def\b{\beta}

\def\d{\partial}

\newcommand{\Cset}{{\,\,{{{^{_{\pmb{\mid}}}}\kern-.45em{\mathrm C}}}}}

\newcommand{\be}{\begin{equation}}
\newcommand{\bea}{\begin{eqnarray}}

\newcommand{\ee}{\end{equation}}
\newcommand{\eea}{\end{eqnarray}}

\newcommand{\pa}{\partial}

\def\pa{\partial}

\begin{document}

\renewcommand{\thefootnote}{\fnsymbol{footnote}}

\makeatletter
\@addtoreset{equation}{section}
\makeatother
\renewcommand{\theequation}{\thesection.\arabic{equation}}

\rightline{}
\rightline{}




\begin{center}
{\LARGE \bf{\sc Euler fluid in 2+1 dimensions  as a gauge theory, and an action for the Euler fluid in any dimensions}}
\end{center} 
 \vspace{1truecm}
\thispagestyle{empty} \centerline{
{\large \bf {\sc Horatiu Nastase${}^{a}$}}\footnote{E-mail address: \Comment{\href{mailto:horatiu.nastase@unesp.br}}
{\tt horatiu.nastase@unesp.br}}
{\bf{\sc and}}
{\large \bf {\sc Jacob Sonnenschein${}^{b,c}$}}\footnote{E-mail address: \Comment{\href{mailto:cobi@tauex.tau.ac.il}}{\tt cobi@tauex.tau.ac.il}}
                                                        }

\vspace{.5cm}


\centerline{{\it ${}^a$Instituto de F\'{i}sica Te\'{o}rica, UNESP-Universidade Estadual Paulista}} 
\centerline{{\it R. Dr. Bento T. Ferraz 271, Bl. II, Sao Paulo 01140-070, SP, Brazil}}
\vspace{.3cm}
\centerline{{\it ${}^b$School of Physics and Astronomy,}}
\centerline{{\it The Raymond and Beverly Sackler Faculty of Exact Sciences, }} 
\centerline{{\it Tel Aviv University, Ramat Aviv 69978, Israel}}
\vspace{1truecm}

\thispagestyle{empty}

\centerline{\sc Abstract}

\vspace{.4truecm}

\begin{center}
\begin{minipage}[c]{380pt}
{\noindent In this paper we parallel the construction of Tong of a gauge theory for  shallow water, by writing a 
gauge theory for the Euler fluid in 2+1 dimensions. We  then extend it to an Euler fluid coupled to electromagnetic background.
We argue that the gauge theory formulation provides a topological  argument  for the quantization of 2+1 dimensional Euler Hopfion solution. 
In the process, we find a (non-gauge) action for the Euler fluid that can be extended to any dimension, including 
the physical 3+1 dimensions. We discuss several aspects of the ABC flow. 
}
\end{minipage}
\end{center}

\vspace{.5cm}

\setcounter{page}{0}
\setcounter{tocdepth}{2}

\newpage

\tableofcontents
\renewcommand{\thefootnote}{\arabic{footnote}}
\setcounter{footnote}{0}

\linespread{1.1}
\parskip 4pt



\section{Introduction}

In a very interesting paper \cite{Tong:2022gpg}, Tong rewrote the shallow water equations, with variables: 2+1 dimensional 
velocity $\vec{u}(x,y,t)$ and height $h(x,y,t)$, in terms of a gauge theory.\footnote{The gauge formulation was  
found to be related to 2 dimensional area preserving diffeomorphisms in \cite{Sheikh-Jabbari:2023eba}.}
In the case of the linearized theory, one obtains a 
Maxwell-Chern-Simons theory, which is known to have boundary chiral modes, that are now identified with the coastal 
Kelvin waves of the shallow water equations, giving them a topological reason. An action for the shallow water and 
the equivalent gauge theory is also obtained. He also suggests that a similar analysis could be made for Euler fluids with 
equation of state $P=C\rho^\gamma$ (so barotropic, $P=P(\rho)$), though that is not done.

In this paper, we show that indeed, similar to the construction in \cite{Tong:2022gpg}, one can  rewrite the 2+1 dimensional 
Euler fluids as a gauge theory. For that purpose we map the magnetic field $B$ to the density of the fluid $\rho$, instead 
of the height function, and find that the magnetic term in the action is linear, and not quadratic in $B$.
The resulting action describes fluids, not necessarily  
 barotropic, such as to include the case of the 2+1 dimensional
fluid Hopfion with constant $\rho$, but nontrivial $P$, considered in 
\cite{Alves:2017ggb,Alves:2017zjt}. 
We  also consider  the case of Euler fluids coupled to 
electromagnetism, as considered by Abanov and Wiegmann \cite{Abanov:2021hio} and for which nontrivial Hopfion solution
were found in \cite{Nastase:2022aps}. The quantization of Hopfion solution will then be related to the quantization of the 
Chern-Simons level.

In the process, we write an action principle for the Euler fluid, which generalizes to any dimension, though only in 2+1 
dimensions is rewritten as a gauge theory.\footnote{The action is in fact equivalent to an action written in \cite{Bistrovic:2002jx}, 
as we became aware after the paper was first posted on the arXiv.} 
Writing an effective action for an Euler fluid was considered before in 
quantum field theory formalisms, starting with \cite{Nickel:2010pr} (based on AdS/CFT holographic arguments), 
and various constructions were attempted in 
\cite{Dubovsky:2011sj,Haehl:2013hoa,Crossley:2015evo,Glorioso:2017fpd,Haehl:2018lcu} (see also other references 
therein), but here the action principle is based simply on the fluid variables. 
We also briefly discuss several aspects of the ABC flow. In particular, we review its  Clebsh formulation, and   
for a special case  of the ABC flow, we map it to electric and magnetic fields,  which we also write in terms of the Bateman 
construction.

The paper is organized as follows. In section 2 we put the Euler fluid in gauge theory form, and we also compare 
the energy-momentum tensors of the fluid and gauge forms. In section 3 we couple the system to electromagnetism, 
and find that this is very natural in the gauge picture. As a further application, we also couple the shallow water
equations to electromagnetism, and write it in the gauge picture. In section 4 we describe the main application 
of the gauge theory picture, describing the Euler fluid Hopfions as  topological modes in the gauge theory. 
In section 5 we first write an action for the Euler fluid in any dimension, and then write a gauge theory form for it, and 
we consider the ABC flow, and present its Clebsh parametrization and a gauge field representation. 
We conclude in section 6.

\section{Euler fluid in gauge theory form}

We consider the 2+1 dimensional Euler (non-dissipative) fluids, with the variables: the flow velocity  $\vec{u}(x,y,t)$, 
the fluid density  $\rho(x,y,t)$, and the pressure $P(x,y,t)$.  
The well-known equations of motion are
\bea
&&\d_t\rho+\vec{\nabla}\cdot(\rho\vec{u})=0\cr
&&\rho\d_t u^i+\d_i P+\rho(\vec{u}\cdot \vec {\nabla})u^i=0\Leftrightarrow \rho\frac{D\vec{u}}{Dt}=-\vec{\nabla}P\;,
\eea
where we have defined the ``covariant derivative", 
\be
\frac{D}{Dt}=\d_t +\vec{u}\cdot\vec{\nabla}.
\ee

We then define the chemical potential $\mu$, as usual, by 
\be
\frac{dP}{\rho}=\frac{d\mu}{m}.
\ee

It would seem like we need a barotropic fluid, $P=P(\rho)$, in order to integrate the relation and find $\mu$, but actually, 
we see that the other case of interest for us, nontrivial $P$ and $\rho$ constant, is also included in it. In general then, 
all we need is to be able to integrate $dP/\rho$. 

We note that in the Appendix of \cite{Tong:2022gpg} it was stated that we could consider the barotropic Euler fluid 
with $P=C\rho^\gamma$, in which case, in the action below (\ref{gaugeaction}), we would replace 
the term with $B\d_0\tilde\mu$ by a term with $B^{\gamma-1}$. However, one of the  reasons we consider our gauge theory 
action for the Euler fluid is that we want to obtain a topological gauge description for the fluid Hopfion  (just like 
Tong obtained a topological gauge description for the coastal Kelvin waves), and for the Hopfion $P$ is a function of 
space, but $\rho$ is constant.

Further, we define $\tilde \mu$ by 
\be
\d_0 \tilde \mu=\frac{\mu}{m}\;,
\ee
which is introduced because we want to make $\tilde \mu$ a variable in the action, and by partial integration its equation 
of motion is $\d_0 $ on what it multiplies equals zero, rather than just what it multiplies equals zero. 

The result of this is that the $\tilde \mu$ equation of motion will enforce $\d_0B=0$, so $\d_0\rho=0$ (density constant 
in time), and by the continuity equation this also means $\vec{\nabla}\cdot (\rho\vec{u})=0$. We could find no way to 
avoid such a restriction, so we obtain a gauge description of the Euler fluid only in this case. 

We then write the gauge theory action for the fluid 
\be
S=\int dt\int d^2x \left[\frac{\vec{E}^2}{2B}-B\d_0 \tilde \mu-\epsilon^{\mu\nu\rho}A_\mu \d_\nu \tilde A_\rho\right]\;,
\label{gaugeaction}
\ee
where $\vec{E},B$ is the gauge field strength that describes the fluid (note that, since $\vec{u},\rho$ are 
observables, they could only be related to fields strengths, not to gauge fields themselves), via the definition
\be
B=\rho\;,\;\;\; E_i=\epsilon_{ij}\rho u^j\;,
\ee
and $\tilde A_\mu$ is an {\em auxiliary} gauge field, defined through the usual Clebsch parametrization as
\footnote{The Clebsch parametrization in the case of fluids was described in 
gauge theory language in \cite{Nair:2011mk}, which also discusses anomalies. In \cite{Nair:2020kjg}, the group theory 
version was used to construct various topological terms.}
\be
\tilde A_\mu=\d_\mu\chi+\b\d_\mu\a\;,\label{Clebsch}
\ee
with $\a,\b,\chi$ real functions that are considered the actual variables (instead of $\tilde A_\mu$). 

The action in terms of these variable reads
\be
S=\int dt\int d^2x \left[\frac{\vec{E}^2}{2B}-B\d_0 \tilde \mu-\epsilon^{\mu\nu\rho}A_\mu \d_\nu(\beta\pa_\rho\alpha)\right]\;,
\label{gaugeactionclebsh}
\ee

Thus, in fact, $\chi$ does not show up in action.

This action is manifestly not Lorentz invariant and instead it is invariant under rotation transformations, and space-time translations.
Due to the CS term the action is also not invariant under parity and time-reversal transformations.
It is obviously also invariant under gauge transformation of $A_\mu\rightarrow A_\mu+ \pa_\mu \lambda$ assuming 
that $\epsilon^{\mu\nu\rho}\lambda\pa_\nu \tilde A_\rho $ vanishes on the boundary of space-time; and separately under
gauge transformations of $\tilde A_\mu$. 

A comparison between this action and the one used in \cite{Tong:2022gpg} \footnote{In \cite{Monteiro:2022wip}, the 
fluid equations of \cite{Nair:2020kjg} were also found to be equivalent to a Chern-Simons action, using a Clebsch parametrization 
for the gauge field.}
reveals the following differences: 
(i) here we have a term linear in $B$ whereas in \cite{Tong:2022gpg} it is quadratic; 
(ii) In the latter there is a term linear in $A_0$  that couples to the Coriolis parameter that our action lacks. 
We have not introduced this term, since it is not present in (exactly) 2+1 dimensional Euler fluids, only in the shallow water 
fluids (2+1 dimensional velocity, but also height of the water, however small). Formally, we could consider 
such a term without worrying 
about where it comes from, just an $f\epsilon^{ij}u^j$ on the right-hand side of $\frac{Du^i}{Dt}$, and then, like for 
\cite{Tong:2022gpg}, this would give an $fA_0$ term in the action. But this is implicitly added in the next section, 
where we couple to external electromagnetic fields $\vec{E}_e,B_e$: we just need to shift the external magnetic field as
\be
\frac{e}{m}B_e\rightarrow \frac{e}{m}B_e+f\;,
\ee
and we obtain the desired term.

Next we consider the equations of motion associated with the variations of $A_\rho$, $\tilde \mu$, $\chi$, $\alpha$ and $\beta$.
We see now that the $\tilde \mu$ equation of motion gives 
\be
\d_0B=\d_0 \rho=0\Rightarrow B={\rm constant}(t)\;,
\ee
instead of just $B=\rho=0$ in the case we wrote simply $\mu/m$ and not $\d_0\tilde \mu$ in the action, 
which would have made it a trivial fluid. As it is though, the action {\em only describes time-independent 
densities $\rho$, but that is enough for our purposes}.\footnote{Note that \cite{Sheikh-Jabbari:2023eba} consider a 
term $BP$ in the Lagrangian for their {\em shallow water} action, but that is necessarily an {\em external} pressure field, 
not a dynamical one as we take here.}

Then, first we see that $\chi$ is a pure gauge, so it has no equation of motion, whereas the equations of motion of 
$\b$ and $\a$ are 
\be
B\dot \a +\epsilon^{ij}E_i\d_j \a=0\;,\;\;\;
B\dot \b +\epsilon^{ij}E_i\d_j \b=0\;,
\ee
and then defining $\tilde E_i$ and $\tilde B$ as the "electric" and "magnetic" fields of $\tilde A_\mu$, such that for instance
\be
\tilde E_i=\dot\b\d_i\a-(\d_i\b)\dot \a\;,
\ee
from the above $\a$ and $\b$ equations of motion we get 
\be
\tilde E_i=\frac{E_i}{B}\tilde B.
\ee

Then the Gauss constraint, i.e., the $A_0$ equation of motion gives
\be
\d_i \left(\frac{E_i}{B}\right)=\tilde B\;,\label{A0eom}
\ee
and, considering that $E_i/B=\epsilon_{ij}u^j$ and that $\d_i(\epsilon^{ij}u_j)\equiv \omega$ is the 2+1 dimensional vorticity,
 replacing in the above equation of motion we get 
\be
\tilde E_i=\epsilon_{ij}u^j \omega.
\ee

The Bianchi identity for the gauge field $A_\mu$ is 
\be
\epsilon^{\mu\nu\rho}\d_\mu(\d_\nu A_\rho)=0\;,
\ee
and it becomes 
\be
\d_0 B +\epsilon^{ij}(\d_i E_j)=0\Rightarrow \d_0 \rho+\d_i(\rho u^i)=0\;,
\ee
i.e., the continuity equation. Since from the equation of motion of $\tilde \mu$, $\d_0\rho=0$, it implies that the action 
describes flows for which
\be
\vec\nabla\cdot(\rho\vec u)=0.
\ee

Finally, the Euler equation appears from the $A_i$ equation of motion of the action, combined with the relation 
$\tilde E_i=\epsilon_{ij}u^j\omega$, obtained previously. Indeed, the $A_i$ equation is 
\be
\d_0\left(\frac{E_i}{B}\right)+\frac{1}{2}\epsilon_{ij}\d_j \left(\frac{\vec{E}^2}{B^2}\right)
+\epsilon_{ij}\d_j \mu/m-\epsilon_{ij}\tilde E_j=0\;,\label{Aieq}
\ee
which translates into 
\be
\d_0(\epsilon_{ij}u^j)+\frac{1}{2}\epsilon_{ij}\d_j(\vec{u}^2)+\epsilon_{ij}\frac{\d_j P}{\rho}=\epsilon_{ij}\tilde E_j=-u_i \omega\;,
\ee
and it is easy to see that, by multiplying the equation by $\epsilon^{ki}$, the left-hand side, minus the term on 
the right-hand side becomes equal to 
\be
-\d_t(u^k)-u_j\d_j u_k-\d_k P=0\;,
\ee
i.e., the Euler equation. 

An interesting observation is that, taking $\d_i$ on the $A_i$ equation of motion (\ref{Aieq}), and using (\ref{A0eom}), 
we obtain 
\be
\d_0\d_i \left(\frac{E_i}{B}\right)=\d_0\tilde B=\epsilon^{ij}\d_i \tilde E_j\;,\label{Mxtilde}
\ee
which is a Maxwell equation in 2+1 dimensions for $\tilde A_\mu$.


\subsection{ The energy momentum tensor}

For the calculation of $T_{\mu\nu}$, the CS term doesn't count (it is independent of the metric), 
so consider the action
\be
\hat S=\int dt\int d^2x \left[\frac{\vec{E}^2}{2B}-B\d_0 \tilde \mu\right].
\ee


We will consider the Belinfante energy-momentum tensor. 
In general, we first couple to a metric, and then write
\be
T_{\mu\nu}=-\frac{2}{\sqrt{-g}}\frac{\delta(\sqrt{-g}{\cal L})}{\delta g^{\mu\nu}}
=-2\frac{\delta {\cal L }}{\delta g^{\mu\nu}}+ g_{\mu\nu}{\cal L}.
\ee

First, we need to understand the fluid system better, to understand what we will be comparing 
against. For a fluid, we have 
\be
T_{\mu\nu}=\rho u^\mu u ^\nu +P(\eta_{\mu\nu}+u_\mu u_\nu)\;,
\ee
where, however, $\rho$ refers to the {\em comoving} (in the local center of mass of the fluid)
total energy density, so is really 
\be
\rho\rightarrow \rho_0=m_0 n_0=m_0 \frac{dN}{dV_0}\;,
\ee
where $m_0$ is the rest mass of the fluid particle, and $dV_0$ is the volume element in the center of 
mass system, related to the moving frame by $dV_0=\gamma dV$. In a nonrelativistic approximation, 
$u^0=\gamma$, so $u_0\simeq -1-v^2/2$, and $u^i =\gamma v^i$, so $u_i \simeq v^i$.
Then, in the moving frame, 
\be
T_{00}=\gamma^2 \rho\simeq \rho+\rho v^2, \;\;\;T_{0i}\simeq -\rho v^i\;,\;\;\;
T_{ij}\simeq \rho  v_i v_j+P\delta_{ij}\;,
\ee
and $\rho$ really refers to $\rho_0$. The energy of the fluid is then 
\be
E=\int dV \gamma^2\rho=\gamma \int dV_0 \rho\simeq \int dV_0 \left[\rho +\rho\frac{v^2}{2}\right].
\ee
However, to this rest+kinetic energy of the fluid particles 
we can add the subleading term of the potential energy $\int dV_0\; P$.

In the nonrelativistic approximation, the total energy contains the term with the rest energy density 
$\rho$, and we integrate over the rest volume, as well as the kinetic energy density. 
With the map $E_i=\epsilon_{ji}\rho v_j$, $B=\rho$, the action $\hat S$ above contains only the 
kinetic energy density (thus omitting the rest mass energy), minus a potential term which, for 
constant $\rho$, is just $PdV_0$. The rest mass energy is neglected since we consider the case 
$\d_t\rho=0$.

The continuity equation and the Euler equation appear, as usual, from the $0$ and $i$ components 
of $\d^\mu T_{\mu\nu}=0$ for the fluid, respectively (after we use the $0$ component in the $i$ 
component also). The difference now is that $\d_t \rho=0$ and, 
instead, we keep just the subleading term with $\d_t (\rho v^2/2)$,  in the 0 component.

To construct $T_{\mu\nu}$ from the action $\hat S$, we couple to the metric as follows. First, 
since $-E_i=F_{0i}$ and $F_{ij}=\epsilon_{ij}B$, $\epsilon_{12}=+1$, 
and we keep these relations even in the 
case of a nontrivial metric, we write 
\bea
\vec{E}^2&=&-F_{0i}F^{0i}=-F_{0i}F_{\mu\nu}g^{0\mu}g^{i\nu}
=-(F_{0i}F_{0j}g^{00}g^{ij}+F_{0j}F_{ik}g^{0i}g^{jk})\cr
2B&=&g^{ik}g^{kl}\epsilon_{jl}F_{ik}\;,
\eea
and we obtain 
\bea
T_{00}&=&-\frac{2}{\sqrt{-g}}\frac{\delta \hat S}
{\delta g^{00}}=\frac{2\vec{E}^2}{2B}-\left(\frac{\vec{E}
^2}{2B}-B\d_0\tilde \mu\right)\cr
&=& \frac{\vec{E}^2}{2B}+B\d_0\tilde \mu\leftrightarrow \rho \frac{\vec{v}^2}{2}+\rho \d_0\tilde \mu.
\eea

The $T_{0i}$ components are obtained from the Lagrangian density term
\be
{\cal L}= -\left[\frac{g^{0j}g^{ik}F_{0i}F_{jk}}{2B}\right ]+... \;,
\ee
so we obtain 
\be
  T_{0i}=
-2\frac{\delta {\cal L }}{\delta g^{0i}}=-\epsilon_{ij} E^j\leftrightarrow -\rho v_i.
\ee

The $T_{ij}$ components are obtained by varying both the $\vec{E}^2$ terms and the $B$ terms 
with respect to $g^{ij}$, so 
\bea
T_{ij}=-\frac{2}{\sqrt{g}}\frac{\delta \hat S}{\delta g^{ij}}&=&-\frac{E_i E_j}{B}+
\frac{\vec{E}^2}{B}\delta_{ij}
+2B\d_0\tilde \mu\delta_{ij}+\left(\frac{\vec{E}^2}{2B}-B\d_0\tilde \mu\right)\delta_{ij}\cr
&=&\frac{-E_iE_j+\vec{E}^2\delta_{ij}}{B}+\left(\frac{\vec{E}^2}{2B}+B\d_0\tilde \mu\right)\delta_{ij}.
\eea

Explicitly, this gives
\bea
T_{11}&=&\frac{E_2^2}{B}+ \left(\frac{\vec{E}^2}{2B}+B\d_0\tilde \mu\right)\leftrightarrow 
\rho\frac{v_1^2}{2}+\left(\frac{\rho \vec{v}^2}{2}+\rho\d_0\tilde \mu\right)\cr
T_{22}&=&\frac{E_1^2}{B}+\left(\frac{\vec{E}^2}{2B}+B\d_0\tilde \mu\right)\leftrightarrow
\rho\frac{v_2^2}{2}+\left(\frac{\rho \vec{v}^2}{2}+\rho\d_0\tilde\mu\right)\cr
T_{12}&=& -\frac{E_1E_2}{B}\leftrightarrow \rho \frac{v_1v_2}{2}.
\eea

We see that we obtain the correct fluid energy $T_{\mu\nu}$, except for an extra term 
$\rho v^2/2 \delta_{ij}$, an extra momentum flux that should be related to the nonrelativistic corrections 
to $\rho$ that we kept: in $\d^0T_{0i}+\d^j T_{ji}=0$, we get an extra $-v_i \d_t(\rho v^2/2)$ 
from the first, and an extra $\d_i (\rho v^2/2)$ from the second, cancelling under the assumption of a 
kinetic energy depending explicitly only on time.

A much more interesting case would have been the viscous Navier-Stokes case. However, it is not easy to generalize
to the presence of the viscous term in the energy-momentum tensor, 
because of the complicated nature of the equations in the gauge theory form. We have not been able to find the 
gauge theory action that corresponds to the Navier-Stokes fluid.

\section{Euler fluid coupled to electromagnetism in gauge theory form}

In this section we consider the Euler fluid equations coupled to external electromagnetism, 
as considered by Abanov and Wiegmann \cite{Abanov:2021hio}\footnote{See also the earlier paper \cite{Monteiro:2014wsa}
for Euler fluids coupled to electromagnetism, and anomalies.}
in 3+1 dimensions, and then by us
\cite{Nastase:2022aps} in 2+1 dimensions, with equations of motion
\bea
&&\d_t\rho+\vec{\nabla}\cdot(\rho\vec{u})=0\cr
&&\rho\d_t u^i+\d_i P+\rho(\vec{u}\cdot \vec {\nabla})u^i=\frac{e}{m}\left(\vec{E}_e+\vec{u}\times \vec{B}_e\right)\;,
\eea
which in 2+1 dimensions gives the Euler equation
\be
\frac{D u^i}{Dt}+\d_i P=\frac{e}{m}\left(E^i_e+B_e\epsilon^{ij}u_j\right)\;,
\ee
where $\vec{E}_e,B_e$ are the true, external (i.e., non-dynamical), electromagnetic fields, and we have put $c=1$ for simplicity.

It is perhaps clear (at least a posteriori, after figuring it out), that the correct way to incorporate the electromagnetic fields 
into the gauge theory action is to add another Chern-Simons term (or rather, BF term) coupling the gauge fields $A_\mu$ 
and $A_\mu^e$ (electromagnetic), so the final action is 
\be
S=\int dt\int d^2x \left[\frac{\vec{E}^2}{2B}-B\d_0 \tilde \mu-\epsilon^{\mu\nu\rho}A_\mu \d_\nu \tilde A_\rho
+\frac{e}{m}\epsilon^{\mu\nu\rho}A_\mu \d_\nu A_\rho^e\right]\;.\label{EulerGem}
\ee

Then the $\a,\b$ equations are unchanged, so is the Bianchi identity for $A_\mu$, giving the continuity equation,
but the Gauss law constraint (equation of motion for $A_0$) gets a new contribution 
from the added term, so is now
\be
\d_i\left(\frac{E_i}{B}\right)-\tilde B +\frac{e}{m}B^e=0\;,
\ee
and gives 
\be
\tilde B =\omega+\frac{e}{m}B^e.
\ee

From the $\a$ and $\b$ equations, we get
\be
\tilde E_i=\frac{E_i}{B}\tilde B =\epsilon_{ij}u^j \left(\omega+\frac{e}{m}B^e\right).
\ee

Then the equation of motion of $A_i$ contains two extra terms, one directly from the coupling of $A_i$ to $E^j_e$, 
and one indirectly, from the fact that now $\tilde E_i$ contains the extra $B^e$ term above, giving 
\be
D_t(\epsilon^{ij} u^j)+\epsilon^{ij}\d_j \frac{\mu}{m}+\frac{e}{m}u^iB^e=\frac{e}{m}\epsilon^{ij}E_j^e\;,
\ee
which is nothing but the Euler equation coupled to external electromagnetism, as we wanted.

The Maxwell equation for $\tilde A_\mu$ in (\ref{Mxtilde}) is then modified as
\be
\d_0\d_i \left(\frac{E_i}{B}\right)=\d_0\left(\tilde B-\frac{e}{m}B^e\right)=\epsilon^{ij}\d_i \left(\tilde E_j-\frac{e}{m} E_j^e\right).
\ee

\subsection{Shallow water equations coupled to electromagnetism, and gauge theory form}

Here we make a small aside, and note that we can also couple the shallow water equations to electromagnetism, 
and write it in gauge theory form, just as above for the Euler case.

Indeed, the shallow water equations coupled to electromagnetism (for velocity $u^i$ and height $h$ of the fluid) are
\bea
&&\d_t h+h\vec{\nabla}\cdot \vec{u}=0\cr
&&\frac{Du^i}{Dt}= f\epsilon^{ij}u^j-g\d_i h+\frac{e}{m}\left(E^i_e+\epsilon_{ij}u^jB_e\right)\;,
\eea
where $f$ is the Coriolis parameter, $g$ is the gravitational acceleration 
and $E^i_e, B_e$ are the true external electric and magnetic fields.

Again, the correct way to introduce the coupling to $E^i_e, B_e$ in the gauge theory formulation is to 
introduce an extra BF term, of the type $AdA_e$, namely to modify the gauge theory action in \cite{Tong:2022gpg}
to (here, as before, we have the Clebsch parametrization $\tilde A_\mu=\d_\mu \chi+\b\d_\mu\a$)
\be
S=\int dt\; d^2x\left[\frac{\vec{E}^2}{2B}-\frac{g}{2}B^2+fA_0-\epsilon^{\mu\nu\rho}A_\mu \d_\nu \tilde A_\rho
+\frac{e}{m}\epsilon^{\mu\nu\rho}A_\mu \d_\nu A_\rho^e\right].
\ee

Indeed, again one finds that only the Gauss constraint (equation of motion for $A_0$) is modified, 
adding a term $\frac{e}{m}B_e$ to $\tilde B$, which becomes 
\be
\tilde B=\omega+f+\frac{e}{m}B_e\;,
\ee
and also implies (from the $\a$ and $\b$ equations of motion)
\be
\tilde E_i =\frac{E_i}{B}\tilde B=\epsilon^{ij}u_j \left(\omega+f+\frac{e}{m}B_e\right)\;,
\ee
as well as the $A_i$ equation of motion, in which we get an extra term from the coupling of $A_i$ to $E^j_e$, 
and an indirect term, via the contribution of $\tilde B$ to $\tilde E_i$ above, 
\be\label{EOMEB}
\d_t\left(\frac{E_i}{B}\right)+\frac{1}{2}\epsilon_{ij}\d_j \left(\frac{\vec{E}^2}{B^2}\right)+g\epsilon_{ij}\d_j B
-\epsilon_{ij}\tilde E_j+\frac{e}{m} \epsilon _{ij}E_e^j=0.
\ee

When substituting the map to the fluid, we get indeed the shallow water equation coupled to electromagnetism, 
\be
D_t(\epsilon^{ij}u^j)+f u^j +g\epsilon^{ij}\d_j h +\frac{e}{m}u^i B_e-\frac{e}{m}\epsilon^{ij}E^j_e=0.
\ee

\section{2+1 dimensional Hopfion as a gauge theory topological mode}

Similar to the way Tong found the coastal Kelvin waves of the shallow water equations as 
chiral boundary modes (topological modes) in the {\em effective} Chern-Simons theory obtained for small 
fluctuations \cite{Tong:2022gpg}, in this section we want to see if the 2+1 dimensional Hopfion solution 
with constant $\rho$ can be similarly understood from topological considerations in the corresponding gauge theory. 

The 2+1 dimensional Hopfion solution was obtained in \cite{Alves:2017ggb,Alves:2017zjt}
by dimensional reduction of the 3+1 
dimensional Hopfion, which is a null ($\vec{u}^2=1$) fluid solution obtained 
from the analogy to 3+1 dimensional electromagnetism with a Hopfion solution (that is also null in electromagnetism).

The 2+1 dimensional Hopfion solution has $\rho=1$, is time independent and has
\bea
P&=&P_\infty -\frac{2}{1+x^2+y^2}\cr
u_x&=&\frac{2y}{1+x^2+y^2}\cr
u_y&=&-\frac{2x}{1+x^2+y^2}\;,\label{HopfionSol}
\eea
which gives the vorticity
\be
\omega=\epsilon^{ij}\d_i u_j=-\frac{4}{(1+x^2+y^2)^2}\;,
\ee
which integrates to $-4\pi$.

\subsection{The winding number in the gauge picture}

We translate the winding number which is the integral over the vorticity to the gauge theory language,
\be
{\cal H}= \int d^2 x\; w(t,x,y) = \int d^2 x\; (\pa_x u_y - \pa_y u_x)= 
\int d^2x\; \pa_i \left (\frac{E^i}{B}\right ) .
\ee

We would like to check if this quantity is conserved in time. For that purpose  we use the equations of motion for $A_i$, 
or more precisely the Maxwell equations (\ref{Mxtilde}) derived from them, and find that
\be
\pa_t  {\cal H}=\int d^2x \; \d_t \tilde B=\int d^2 x\;\epsilon^{ij}\pa_i\tilde E_j
=\int d^2x \;\epsilon^{ij}\d_t(\pa_i \beta \pa_j\alpha-\pa_i \alpha \pa_j\beta).
\ee 

Thus, the winding ${\cal H}$ is conserved in time, since it is a total derivative, so a boundary term, that can be put to zero.
We also note that, in the more general case of coupling to an electromagnetic field $B_e, \vec{E}_e$, if we have a 
constant $B_e$ and vanishing $\vec{E}_e$, the same result applies.


\subsection{ Linearized fluctuations}

Again following the logic in \cite{Tong:2022gpg}, we write a linearized theory by expanding around a 
background with $\hat \a, \hat \b,\rho_0,\hat A_\mu$, but with a small (perturbation) velocity $u^i=\delta u^i$
and $\omega=\delta \omega$, so $E_i/B= \epsilon^{ij}u_j$, as
\bea
&&\a=\hat \a+q\;,\;\; \b=\hat \b+p\cr
&&\d_1\hat\b\d_2\hat \a-\d_2\hat\b\d_1\hat\a\equiv k\Rightarrow \d_j\hat \b\d_l \hat \a-\d_l \hat \b\d_j\hat \a=k\epsilon_{jl}\cr
&&B=\rho_0+\delta b\;, E_i=\delta e_i\cr
&&A_\mu=\hat A_\mu+\delta A_\mu\;,\;\;\; \hat A_0=0\cr
&\Rightarrow&\d_1\hat A_2-\d_2\hat A_1=\rho_0\;,
\eea
and writing simply $A_\mu$ for $\delta A_\mu$, so $B$ for $\delta b=\delta \rho$ and $E_i$ for $\delta e_i=\rho_0 \epsilon
^{ij}u_j$.
Note that we have defined
\be
\tilde B =\d_1\hat \b \d_2\hat \a-\d_2\hat \b\d_1\hat \a\equiv k\;,
\ee
but the Gauss constraint of the action (\ref{EulerGem}) is, in the above notation for linearization,
\be
k\equiv \tilde B=\omega +\frac{e}{m}B_e(+f)\;,
\ee
where we have put an $f$ which is just a shift of $B_e$,
but we have assumed that $\omega$ is a perturbation, $\omega=\delta \omega$,
so the only way to assume $k$ is not (as we will see that 
we need) is to say that $B_e$ or $f$ are large. That, in turn, means that the Hopfion solution (\ref{HopfionSol}) is 
not valid, as it was derived at $B_e=f=0$.  Nevertheless, we would still like to see if there is a quantization 
condition possible in this (as of yet not considered) case for the Hopfion.

Then the action becomes 
\be
S=\int dt\int d^2x \left[\frac{\vec{E}^2}{2\rho_0}-\rho_0\d_0\tilde \mu-\delta b\d_0\tilde \mu-\rho_0 p\dot q
+\epsilon^{ij}E_i(q\d_j \hat \b-p\d_j\hat \a)\right].
\ee

The equations of motion for $p,q$ (coming originally from $\tilde A_\mu$) are 
\be
\rho_0\dot q=-\epsilon^{ij}E_i\d_j\hat \a\;,\;\;\;
\rho_0\dot p=-\epsilon^{ij}E_i\d_j \hat \b\;,
\ee
the equation of motion for $A_i$ is 
\be
\dot E_i=-\rho_0 \epsilon_{ij}\d_j \d_0\tilde \mu-\rho_0\epsilon_{ij}(\d_j\hat\b \dot q -\d_j\hat \a \dot p)\;,
\ee
and the equation for $A_0$ (Gauss constraint, for the $A_0=0$ gauge) is 
\be
\d_i E_i =\rho_0\epsilon_{ij}(\d_i\hat \b\d_j q-\d_i\hat \a \d_jp).
\ee

Replacing $\rho_0 \dot q$ and $\rho_0 \dot p$ from the equations for $q$ and $p$ into the equation for $A_i$, 
and using $\d_j\hat \b\d_l \hat \a-\d_l \hat \b\d_j\hat \a=k\epsilon_{jl}$, we obtain 
\be
\dot E_i =\epsilon_{ij}[kE_j-\rho_0\d_j\d_0\tilde \mu].\label{Eeq}
\ee

But also using the solutions 
\be
\rho q=-\epsilon^{ij}A_i\d_j\hat \a\;,\;\;
\rho p=-\epsilon^{ij}A_i \d_j\hat \b
\ee
of the equations for $p$ and $q$, and using $\d_j\hat \b\d_l \hat \a-\d_l \hat \b\d_j\hat \a=k\epsilon_{jl}$, we obtain 
that the Gauss constraint in the $A_0=0$ gauge becomes
\be
\d_i E_i=k B\;,
\ee
and the effective action for the linearized fluctuations becomes 
\be
S=\int dt\int d^2x \left[\frac{\dot A_i^2}{2\rho_0}-B\d_0\tilde \mu+\frac{k}{2\rho_0}\epsilon^{ij}A_i\dot A_j\right]\;,
\ee
whose $A_i$ equation is the same as (\ref{Eeq}), so that the full action (including $A_0$) is 
\be
S=\frac{1}{2\rho_0}\int dt\int d^2x \left[E^2-2B\rho_0\d_0\tilde \mu -k\epsilon^{\mu\nu\rho}A_\mu \d_\nu 
A_\rho\right]\;,
\ee
which contains now a true CS term (not a BF term like the full nonlinear theory)!

Note that in fluid variables, the Gauss constraint $\d_i E_i=kB$ becomes, at this linearized level,
\be
\d_i E_i\simeq \rho_0\epsilon^{ij}\d_i u_j\equiv \rho_0\omega =k\delta\rho\Rightarrow \omega=\frac{k}{\rho_0}\delta \rho.
\ee

But then, from the quantization of the coefficient of the CS term, we obtain $(k/\rho_0)\in 2\pi \mathbb{Z}$, so 
$\omega/\delta \rho\in2\pi \mathbb{Z}$, 
which is the topological reason we wanted for the quantization of vorticity $\omega$ when we perturb the 
constant density $\rho_0$ by $\delta \rho$, in the background of a constant density $\rho_0=1$ and a 
constant external magnetic field $B_e$ (or Coriolis term $f$).

Finally, that means that the 2+1 dimensional fluid Hopfion with quantized vorticity appears because of the 
existence of quantized linearized modes in the gauge theory action.

\section{An action for an Euler fluid in any dimension and a 2+1 dimensional gauge action}

Like in the Appendix of  \cite{Tong:2022gpg} (for the shallow water case $h,\vec{u}$), 
one can start with an action for the 2+1 dimensional Euler fluid (with variables $\rho,\vec{u},P$), and 
derive the same gauge theory action directly from it. However, now we will focus on the 
Euler fluid action, and generalize it to higher dimensions. 

We start with the Euler fluid action 
\be
{\cal L}=\frac{\rho \vec{u}^2}{2}-\rho\d_0\tilde\mu +\phi\left(\frac{D\rho}{Dt}+\rho \vec{\nabla}\cdot \vec{u}\right)-\rho
\b_a\frac{D\a^a}{Dt}\;,\label{Euleraction}
\ee
where $\phi$ is a Lagrange multiplier and $\a^a,\b_a$ are variables from a Clebsch parametrization of the velocity $\vec{u}$, 
as we will see.

In this action, the first two terms are obvious.
Indeed, the first term is just the kinetic term of the fluid, the second is $=-\rho\mu/m$, where $-\d\mu/m=-dP/\rho$, 
so the term is just (at least if $\rho$ is constant) $P$, so for the Lagrangian $L$ (not the Lagrangian density ${\cal L}$) this would be 
$-\int P dV$, giving a potential energy. The third term just enforces the conservation of the energy density $\rho$, 
\be
\frac{D\rho}{Dt}+\rho \vec{\nabla}\cdot \vec{u}=\frac{\d \rho}{\d t}+\vec{\nabla}(\rho\vec{u})=0\;,
\ee
with the Lagrange multiplier $\phi$. The last term is more unusual, but is the exact analog of the corresponding term for 
the shallow water action in  \cite{Tong:2022gpg}. 

We then have equations of motion for the variables $\rho,\vec{u}, \phi,\a^a,\b_a$.

We want to obtain the Euler equation from the Lagrangian, which in our definitions is 
\be
\frac{D\vec{u}}{Dt}=-\frac{\vec{\nabla}P}{\rho}=-\vec{\nabla}\mu=-\vec{\nabla}\d_0\tilde \mu.
\ee

1) The equation of motion for $\rho$ gives 
\be
\frac{\vec{u}^2}{2}-\d_0\tilde \mu-\frac{D\phi}{Dt}=0\Rightarrow
\frac{D\phi}{Dt}=\frac{\vec{u}^2}{2}-\frac{\d_t P}{\rho}.
\ee

2) The $\phi$ equation of motion gives, as we said, the conservation of $\rho$ (the continuity equation),
\be
\d_t\rho+\vec{\nabla}\cdot(\rho\vec{u})=0.
\ee

3) The $\a^a$ equation of motion gives 
\be
\d_t(\rho\b_a)+\vec{\nabla}\cdot(\rho \b_a \vec{u})=0\;,
\ee
but using the conservation of $\rho$ above, this gives the conservation of $\b_a$,
\be
\frac{D\b_a}{Dt}=\d_t\b_a +\vec{\nabla}\cdot(\b_a\vec{u})=0.
\ee

4) The $\b_a$ equation of motion gives the conservation of $\a^a$,
\be
\frac{D\a^a}{Dt}=0.
\ee

5) The equation of motion of $\vec{u}$ gives the form of $\vec{u}$ in a Clebsch-like parametrization, 
\be
\vec{u}=\vec{\nabla}\phi+\b_a\vec{\nabla}\a^a.\label{Clebschaction}
\ee

Now the Euler equation is obtained as a combination of the $\vec{u}, \a^a,\b_a $ and $\rho$ equations.

We first use the $\vec{u}$ equation of motion (the resulting Clebsch-like parametrization) to calculate first 
\bea
\frac{Du^i}{Dt}&=& \d_t u^i+\vec{\nabla}\cdot \vec{\nabla} u^i\cr
&=& \d_t \d_i \phi+\d_t(\b_a\d_i \a^a)+u^j\d_j (\d_i \phi+\b_a\d_i \a^a)\cr
&=&\d_i \d_t\phi+(u^j\d_j)\d_i\phi\cr
&&+(\d_t\b_a)(\d_i\a^a)+(u^j\d_j \b_a)(\d_i \a^a)\cr
&&+\b_a \d_i \d_t \a^a+\b_a (u^j\d_j \d_i \a^a).
\eea

In the last expression, we use the $\rho$ equation of motion $D\phi/Dt=0$ in the first line, the second line vanishes due 
to the $\a^a$ equation of motion (that used also the $\phi$ equation, the $\rho$ conservation, as we saw) $D\b_a/Dt=0$, 
and in the third line we use the $\b_a$ equation of motion $D\a^a/Dt=0$, to obtain 
\bea
\frac{Du^i}{Dt}&=& -\b_a (\d^i u^j)\d_j \a^a-(\d^i u^j)\d_j \phi +\d^i \frac{\vec{u}^2}{2} -\d^i \d_t \tilde \mu\cr
&=&-\d^i\d_t \tilde\mu\;,
\eea
where in the last equality we have used again the equation of motion of $u_i$, that $\d_j\phi+\b_a\d_j\a^a=u_j$.

We see that we finally obtained the Euler equation, as we wanted. 

The above action is trivially generalized to 3+1 dimensions, as nothing in it, 
not the action, and not the above derivation, depends on dimension.

The equation of motion for $\vec{u}$ in the above action gives the Clebsch parametrization for the velocity (\ref{Clebschaction}), 
which is true in any dimension. In particular, it has been used in 3+1 dimensions, as we note in the next subsection.

We can also integrate by parts $D\rho/Dt=\d_t \rho+\vec{u}\cdot \vec{\nabla}\rho$, in order to 
obtain $\rho$ as a common factor of the action, obtaining the Euler action in general dimension
\footnote{This action has essentially been used by \cite{Bistrovic:2002jx}, as we became aware after the first version of this
paper was posted on the arXiv. Indeed, in \cite{Bistrovic:2002jx}, one uses the Lagrangian
\be
{\cal L}=-j^\mu a_\mu +\frac{1}{2}\rho \vec{v}^2-V\;,
\ee
where $V$ is some potential giving a pressure (or force), and 
\be
j^\mu=(c\rho,\rho\vec{v})\;,\;\;\;
a_\mu=\d_\mu\theta+\a^i\d_\mu\b^i\;,
\ee
so the auxiliary gauge field $a_\mu$ is given in a Clebsch parametrization.

Writing explicitly, we have 
\be
{\cal L}=\rho\left[\frac{\vec{v}^2}{2}-\frac{V}{\rho}-\frac{D\theta}{Dt}-\a^i\frac{D\b^i}{Dt}\right].
\ee

We see that it is basically our action, since $-\int V/\rho=\int \d_0\tilde\mu$. It is not clear to us whether \cite{Bistrovic:2002jx}
meant for it to be used in any dimension (since in fact, they did not have indices $i$ on $\a^i, \b^i$, but rather a single set).

Note that the extension of \cite{Bistrovic:2002jx}  to cases with spin was done in \cite{Karabali:2014vla}.}

\be
{\cal L}=\rho\left[\frac{\vec{u}^2}{2}-\d_0\tilde\mu -\frac{D\phi}{Dt}-\b_a \frac{D\a^a}{Dt}\right].
\ee

The transition to a gauge theory action works, however, only in 2+1 dimensions. 

We introduce the Lorentz covariant parametrization $C_\mu\equiv \d_\mu \phi$, and consider now $C_\mu$ as 
the independent field, and impose the parametrization with the new Lagrange multiplier = gauge field $A_\mu$, 
via a new term in ${\cal L}$, 
\be
{\cal L}\rightarrow \rho\left[\frac{\vec{u}^2}{2}-\d_0\tilde\mu -(C_0+u^i C_i)-\b_a \frac{D\a^a}{Dt}\right]
+\epsilon^{\mu\nu\rho}A_\mu \d_\nu C_\rho.
\ee

Then the equation of motion for $C_0$ gives
\be
B\equiv \d_1 A_2-\d_2 A_1=\rho\;,
\ee
the equation of motion for $C_i$ gives
\be
E_i \equiv -(\d_0 A_i-\d_i A_0) =\epsilon_{ij}\rho u ^j\;,
\ee
and the rest are as before. 

Imposing these equations of motion amounts to dropping the terms $-\rho(C_0+u^i C_i)$ and $\epsilon^{\mu\nu\rho}A_\mu
\d_\nu C_\rho$ from the action, and replacing $(\rho, u^i)$ with $(B, E_i)$ according to the above map, which gives
\bea
\frac{\rho \vec{u}^2}{2}&=& \frac{\vec{E}^2}{2B}\;,\cr
\b_a \frac{D\a^a}{Dt}&=& (\rho) \b_a \d_t \a^a+(\rho u^j)\b_a \d_j \a^a\cr
&=& (\d_1 A_2-\d_2 A_1)\b_a \d_t \a^a-(\d_t A_i -\d_i A_0)\epsilon^{ik}\b_a \d_j \a^a\cr
&=& \epsilon^{\mu\nu\rho}(\d_\mu A_\nu) \b_a \d_\rho \a^a.
\eea

Thus the action becomes 
\be
{\cal L}=\frac{\vec{E}^2}{2B}-B\d_0\tilde \mu-\epsilon^{\mu\nu\rho}(\d_\mu A_\nu) \b_a \d_\rho \a^a\;,
\ee
which is just the action (\ref{gaugeaction}) with the Clebsch parametrization (\ref{Clebsch}).

\subsection{The ABC flow, in Clebsh and gauge field representations}

As an example of the Clebsch parametrization for Euler flow
in the case of 3+1 dimensions, 
the standard ABC flow\footnote{ We thank P. Wiegmann for bringing the ABC flow to our attention.} 
solution of the Euler equations in 3+1 dimensions, 
\bea
u_x&=&\dot x=b\sin y -c \cos z\cr
u_y&=&\dot y =c\sin z -a \cos x\cr
u_z&=& \dot z=a\sin x-b \cos y\;,
\eea
can be written in the Clebsch parametrization,
\be
\vec{u}=\vec{\nabla}\phi+\b_1\vec{\nabla}\a^1+\b_2\vec{\nabla}\a^2\;,
\ee
where \cite{2017PhRvL.119x4501Y}
\bea
\phi& =& z(a\sin x-b \cos y)\cr
\b_1&=& b\sin y -c\cos z -az \cos x\cr
\b_2&=& c\sin z -a\cos x-bz \sin y\cr
\a_1&=& x\cr
\a_2&=& y.
\eea

Can one map the ABC flow to a gauge field configuration? In \cite{Alves:2017ggb,Alves:2017zjt} we proposed a map 
between fluid and gauge dynamics by mapping the corresponding components of the energy momentum tensor. 
This map to ordinary Maxwell theory in  3+1 dimensions is valid for $\vec{u}^2=1$ (null motion, $u^\mu u_\mu=0$)
and takes the form (here $a=1,2,3$)
\be
T_{00}= \rho\leftrightarrow \frac12(\vec{E}^2+\vec{B}^2) \qquad T_{a0}= \rho u_a \leftrightarrow [\vec{E}\times \vec{B}]_i.
\ee

We start with the degenerate ABC flow where $b=c=0, a=1$. In this case the flow vector is given by
\be
\vec v =(0, -\cos x, \sin x)\;,\qquad \rho =1.
\ee

It is easy to check that in this case the corresponding  electric and magnetic fields are given by 
\be
\vec E= (1,0,0)\;, \qquad \vec B =- (0, \sin x,\cos x)=\vec{\nabla}\times \vec{A}\Rightarrow 
\vec{A}=-(0,\sin x,\cos x)=\vec{B}.\label{EBspecial}
\ee

It is interesting to note that  the dual (electromagnetic) of this  special ABC flow is 
a ``null configuration''\cite{Hoyos:2015bxa} obeying
\be
\vec E\cdot\vec B=0\;, \qquad \vec{E}^2=\vec{B}^2.
\ee

As such the electromagnetic fields admit  conserved nontrivial helicities, in particular the nonzero magnetic-magnetic 
helicity,
\be
{\cal H}_{\rm mm} =\int d^3 x \vec A\cdot\vec B=  \int d^3 x( \cos ^2 x + \sin^2 x) =\int d^3 x.
\ee 

Note however that this magnetic field $\vec B$ does not obey the (free, vacuum) Maxwell equations, since 
\be
\vec \nabla \times \vec B= -(0,\sin x, \cos x)=\vec B\;,
\ee
so at most it can be interpreted as being sourced by a current $\vec J=-(0,\sin x, \cos x)=\vec B$ (but the 
electromagnetic helicities are usually defined for Maxwell fields in vacuum).

Next we can express the electric and magnetic fields in terms of two complex scalar field $\alpha$ and $\beta$ 
using the  Bateman formulation\cite{Hoyos:2015bxa}, namely
\be
\vec F= \vec E + i \vec B =\vec \nabla \alpha\times \vec \nabla \beta.
\ee

The complex scalar fields that yield the electric and magnetic fields given above are 
\be
\alpha=y- i \cos x\;, \qquad \beta =z + i\sin x.
\ee

In the Bateman formulation, half of the (free, vacuum) Maxwell's equations are automatic, $\vec{\nabla}\cdot \vec{F}=0$, 
while the other half are 
\be
\vec{\nabla}\times \vec{F}=i\d_t \vec{F}\Rightarrow i(\d_t\a\vec{\nabla}\b-\d_t\b\vec{\nabla}\a)=\vec{F}=\vec{\nabla}\a\times
\vec{\nabla}\b\;,
\ee
and these are not satisfied, since $\d_t \a=\d_t\b=0$, statement equivalent to 
\be
\vec{\nabla}\times \vec{B}-\d_t \vec{E}\neq 0\;,\vec{\nabla}\times\vec{E}+\d_t\vec{B}\neq 0\;,
\ee
which we already noted happens in this case ($\vec{\nabla}\times \vec{B}-\d_t \vec{E}=\vec{B}$ now). 

Can we map a more general ABC flow to electromagnetism? The answer turns out to be no, as we see now. 
Basically, the problem is that the map is only valid for $\vec{u}^2=1$, which for the ABC flow becomes
\be
a^2+b^2+c^2-2bc\sin y \cos z -2ac \sin y \cos x -2ab \sin x \cos y=1\;,
\ee
which we can easily see that it cannot be solved for arbitrary $x,y,z$ in the case of nonzero $a,b,c$.
In fact, in order to have a solution at arbitrary $x,y,z$, we need to have at least two of $a,b,c$ vanish, so that 
$ab=bc=ac=0$. 

We also note that the special case considered, with $b=c=0$, obeys the two-dimensional relations 
(here $i=1,2$)
\be
u_i=\epsilon_{ij}\frac{E_i}{|\vec{E}|}\;,\;\; \rho=|\vec{E}|^2\;,\;\; B=|\vec{E}|\;,
\ee
where $B=B_z$ and $E_i=(E_x,E_y)$ correspond to the formulation of the 2+1 dimensional electromagnetism 
(though, of course, it is not quite 2+1 dimensional, since all are functions of the third coordinate, $z$). 

Finally, in this case, also the space-space components of the energy-momentum tensor match between the fluid and 
electromagnetism, since (replacing $B_a=B\delta_{a3}=|\vec{E}|\delta_{a3}$ in the electromagnetic $T_{ab}$)
\be
T_{ab}=\rho u_a u_b =|\vec{E}^2|\delta_{ab}-E_a E_b -|\vec{E}^2|\delta_{a3}\delta_{b3}\;,
\ee
which we can check separately for the cases $a=3=z$ and $a,b\neq 3=z$.




\section{Conclusions and discussion}

In this paper, we have written a gauge theory for the Euler fluids in 2+1 dimensions, with or without a 
coupling to electromagnetism. Using it, we were able to obtain the 
quantization of vorticity of 2+1 dimensional Hopfion solutions from the quantization of the level of Chern-Simons. 
As a small aside, we have also coupled the shallow water equations to electromagnetism, and wrote the 
resulting equations in gauge theory form.
We have also written an action for the Euler fluid in any dimension, using the Clebsch parametrization 
for the velocity.\footnote{Though an equivalent form was written before, in \cite{Bistrovic:2002jx}.} 

There are several open questions that deserve further research. In particular: 
\begin{itemize}
\item
An important challenge is the quest for gauge theory formulation  that corresponds to the Navier-Stokes  fluid in 2+1 dimensions.
\item
We derived the action for an Euler Fluid coupled to an electromagnetic background. 
A not less interesting  phenomenon is the incorporation of the electromagnetic interactions of the fluid itself.
\item
The action we derived for the Euler fluid in any dimensions includes a Lagrange multiplier term for the continuity equation. 
It would be more elegant to have an action that yields that equation not due to a Lagrange multiplier.
\item
The study of the interplay between fluid flows and topologically non-trivial electromagnetic solutions has been touched 
upon in  this note. There several additional  questions about it. In particular the use of special conformal transformation 
to derive novel solutions following \cite{Hoyos:2015bxa}.

\end{itemize}

\section*{Acknowledgements}

The work of HN is supported in part by  CNPq grant 301491/2019-4 and FAPESP grants 2019/21281-4 
and 2019/13231-7. HN would also like to thank the ICTP-SAIFR for their support through FAPESP grant 2016/01343-7.
The work of J.S was supported  by a center of excellence of the Israel Science Foundation (grant number
2289/18).

\bibliography{EulerGauge}
\bibliographystyle{utphys}

\end{document}